\documentclass[3p]{elsarticle}
\usepackage[utf8]{inputenc} 

\usepackage{amsmath, amsfonts, amssymb, amsthm} 
\usepackage{bm}

\usepackage{xcolor}

\usepackage{lineno}
\usepackage{hyperref}
\usepackage[normalem]{ulem}

\newcommand{\pd}[2]{\frac{\partial #1}{ \partial #2}}   

\DeclareMathOperator{\tr}{tr}

\def\R{\mbox{\(\mathbb{R}\)}}

\usepackage{setspace}
\journal{Acta Biomaterialia}

\begin{document}

\begin{frontmatter}

\title{A Data-Driven Computational Model for Engineered Cardiac Microtissues}

\author[umichbme]{Javiera Jilberto}
\ead{jilberto@umich.edu}

\author[umichbme]{Samuel J. DePalma}
\ead{samdep@umich.edu}

\author[umichbme]{Jason Lo}
\ead{lojason@umich.edu}

\author[bumeche]{Hiba Kobeissi}
\ead{hibakob@bu.edu}

\author[umichbme]{Lani Quach}
\ead{laniq@umich.edu}

\author[bumeche]{Emma Lejeune}
\ead{elejeune@bu.edu}

\author[umichbme]{Brendon M. Baker}
\ead{bambren@umich.edu}

\author[umichbme,umichcard,kingsbme]{David Nordsletten~\texorpdfstring{\corref{CA}}{}} 
\cortext[CA]{Corresponding author}
\ead{nordslet@umich.edu}

\address[umichbme]{Department of Biomedical Engineering, University of Michigan, MI, USA}
\address[umichcard]{Department of Cardiac Surgery, University of Michigan, MI, USA}
\address[kingsbme]{Department of Biomedical Engineering, School of Imaging Sciences and Biomedical Engineering, King's College London, London, UK}
\address[bumeche]{Department of Mechanical Engineering, Boston University, MA, USA}
        
\begin{abstract}

Engineered heart tissues (EHTs) present a potential solution to some of the current challenges in the treatment of heart disease; however, the development of mature, adult-like cardiac tissues remains elusive. Mechanical stimuli have been observed to improve whole-tissue function and cardiomyocyte (CM) maturation, although our ability to fully utilize these mechanisms is hampered, in part, by our incomplete understanding of the mechanobiology of EHTs. In this work, we leverage the experimental data produced by a mechanically tunable experimental setup to generate tissue-specific computational models of EHTs. Using imaging and functional data, our modeling pipeline generates models with tissue-specific ECM and myofibril structure, allowing us to estimate CM active stress. We use this experimental and modeling pipeline to study different mechanical environments, where we contrast the force output of the tissue with the computed active stress of CMs. We show that the significant differences in measured experimental forces can largely be explained by the levels of myofibril formation achieved by the CMs in the distinct mechanical environments, with active stress showing more muted variations across conditions. The presented model also enables us to dissect the relative contributions of myofibrils and extracellular matrix to tissue force output, a task difficult to address experimentally. These results highlight the importance of tissue-specific modeling to augment EHT experiments, providing deeper insights into the mechanobiology driving EHT function.

\end{abstract}


\end{frontmatter}

\section{Introduction}

The development of engineered heart tissues (EHTs) for use in regenerative therapies, drug testing, and disease modeling has the potential to improve the life expectancy of millions of people that suffer from cardiac disease \cite{Tsao2022}. Current state-of-the-art EHTs are manufactured using cardiomyocytes (CMs) derived from induced pluripotent stem cells (iPSCs) and scaffolds that mimic the structure and mechanics of native cardiac tissue \cite{Zhao2020}. However, iPSC-CM maturation in current EHT platforms remains a challenge \cite{Tenreiro2021}. This is evidenced by the lack of hallmark attributes of mature CMs, such as myofibril alignment, protein expression, calcium handling, and electrophysiological response, among others \cite{Karbassi2020}. Biophysical stimuli, such as electrical pacing \cite{Ruan2016, RonaldsonBouchard2018} and mechanical loading \cite{DePalma2021, Leonard2018, Bliley2021}, have been shown to enhance iPSC-CM maturation in different EHT platforms. However, the underlying mechanisms driving these observations remain incompletely understood \cite{Tu2018}, preventing scientists from efficiently optimizing the application of these techniques. 


Several \textit{in-vitro} studies have attempted to elucidate the impact of different mechanical microenvironmental perturbations on the maturity of iPSC-CMs. For example, Leonard et al. \citep{Leonard2018} showed that increasing the resistance against which the EHTs contract increases force generation. Similarly, Bliley et al. \citep{Bliley2021} showed that EHTs grown under passive stretch evolve into more mature tissues. Furthermore, other studies have shown that culturing CMs on hydrogels with a modulus similar to that of the healthy adult myocardium enhances electro-mechanical activity \citep{Koerner2021, Ribeiro2015}. DePalma et al. expanded upon this, showing that iPSC-CMs adapt their contractile behavior in response to ECM mechanics on synthetic, fibrous matrices that better mimic the anisotropic mechanics of the cardiac ECM \cite{DePalma2021}. This and other work by Allen et al. \cite{Allen2019} also showed that increasing the anisotropy of the fibrous matrix results in more aligned myofibrils. These studies highlight the impact of different mechanical cues on EHT formation and function. However, the inherent variability of EHTs – from their formation, maturation, ECM properties and structure, myofibril formation, etc. – cloud our interpretation of each mechanical parameter's relevance and the underlying mechanobiological drivers.

A way of tackling this complex task and providing insight into the multifaceted differences in EHTs is through biomechanical modeling. Biomechanical models have been used to understand the mechanics of \textit{in-vitro} systems, helping decipher the mechanobiology behind the alignment of cells \cite{Thavandiran2013,Das2022}, the biomechanics of microtissue failure due to necking \cite{Wang2013}, the cell force transmission through fibrous substrates \cite{Robison2016, Wang2014}, and mechanics at cell-to-cell junctions \cite{Ronan2015}. These models enable the examination of experimental conditions – as well as exploration through \textit{in silico} testing whereby scenarios that would be virtually impossible to construct experimentally can be studied. Biomechanical models – particularly at the whole-organ level – have further enabled the integration and assimilation of imaging and patient data \cite{Miller2021, Asner2017, Bracamonte2022, Balaban2018}, providing pipelines for generating \emph{patient-specific models} describing cardiac function. These pipelines enable the integration of realistic structure/function and provide a platform for understanding the significance of these factors on localized mechanics, such as strain or stress.


To bypass many of the uncertainties associated with EHTs and delve into the underlying impact of structure and function, we propose the integration of a novel EHT platform with \emph{tissue-specific} computational models. In this study, we leverage the experimental data obtained in the fibroTUG platform \cite{DePalma2023}, which uses electrospinning techniques to generate synthetic, fibrous matrices with defined alignment and material characteristics. Imaging provides detailed information on the ECM and myofiber architecture on a tissue-by-tissue basis. Using image-based modeling and data assimilation, we create \textit{in-silico} twins of individual fibroTUGs and explore how different factors of the ECM and myofibril structure impact localized cellular function and the resultant force measures commonly reported in the literature. Through this integrated experimental-computational approach, we observe that the resultant forces of EHTs can be substantially biased by ECM alignment and mechanobiological factors driving myofibril formation and function.

The paper is structured as follows: in the methods section, we detail the process of creating the computational model from the experimental data. In the results section, we validate our model by comparing the simulation results to image-based algorithms and use experimental and non-experimental conditions to analyze the role of the different mechanical variables in the iPSC-CMs stress and force output relationship in EHTs. We follow with a discussion of the model results and how we can use computational modeling approaches to explore the mechanobiology of these tissues.

\section{Materials and Methods}

\subsection{Experimental Setup} \label{sec:methods:expsetup}

\begin{figure}[ht]
    \hfuzz=5.002pt 
    \centering
    \includegraphics[width=\textwidth]{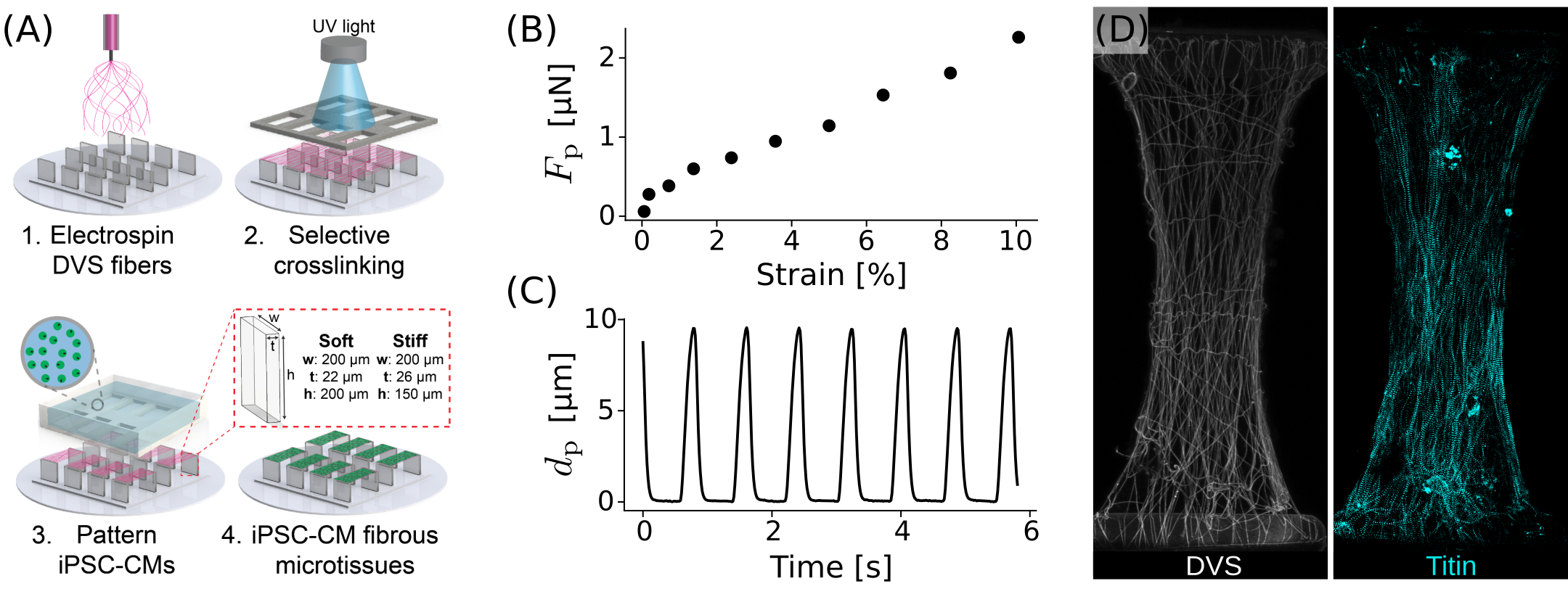}
    \caption{Experimental setup and data output. (A) Diagram of FibroTUG manufacturing process showing the electrospun fibers, the selective crosslinking, and the cell seeding process. (B) Force vs. strain curves obtained by performing an indentation test in the fibers only. (C) Post displacement traces obtained by tracking the post position in videos of the contractile EHTs. (D) Immunofluorescence images of the DVS fibers and the titin protein that indicates the location of Z-discs. }
    \label{fig:expsetup}
\end{figure}

FibroTUG microtissues were fabricated as described in DePalma et. al (2023) \cite{DePalma2023} (see also Fig. \ref{fig:expsetup}A). Briefly, dextran vinyl sulfone (DVS) fibers were electrospun onto an array of polydimethylsiloxane (PDMS) posts attached to a rotating mandrel. The bending stiffness of the posts $k_{\rm p}$  can be tuned by altering the geometry of the posts and is measured experimentally, as described in DePalma et al. (2023) \cite{DePalma2023}. By varying the rotational velocity of the mandrel, the alignment of the fibers can also be controlled, resulting in aligned fibrous matrices that exhibit high anisotropy or random matrices that are more locally isotropic \cite{DePalma2021}. The DVS fibers present between posts were stabilized by exposing them to UV light. Upon hydration, the unstabilized fibers dissolve, leaving only the fibers suspended between two posts. Secondary crosslinking in solutions with varying concentrations of photoinitiator (LAP) resulted in matrices of varying stiffness (higher LAP results in stiffer matrices). The stiffness of the matrix was characterized by indenting the matrices incrementally, measuring the resulting post-deflection (and thus the post-force $F_{\rm p}$), and then calculating the global strain of the matrix (Fig. \ref{fig:expsetup}B). Further, images of the indented DVS fibers were taken to record their organization and pair it with their force response. This setup allows us to tune and control the post stiffness (soft, $k_{\rm p}=0.41$ N/m; stiff, $k_{\rm p}=1.2$ N/m), fiber alignment (aligned or random), and fibrous matrix stiffness (soft, LAP$=0.1$ mg/mL; stiff, LAP$=5.0$ mg/mL), parameters that determine the mechanical environment where the iPSC-CMs develop \cite{DePalma2023}. While many permutations are possible, in this paper, we studied the following permutations: soft fibers/soft posts, stiff fibers/soft posts, and soft fibers/stiff posts for both aligned and random matrices, leading to a total of six conditions.

After defining matrix conditions, purified cultures of iPSC-CMs were seeded onto the fibroTUG and cultured in this environment for seven days. On day 7, time-lapse videos of the microtissue’s spontaneous contractions were acquired (see Video S1). These videos were processed to obtain post-displacement curves as seen in Fig. \ref{fig:expsetup}C. Finally, immunofluorescence staining is used to image cell nuclei, titin, and the DVS fibers (see Fig. \ref{fig:expsetup}D).


\subsection{Image Processing} \label{sec:methods:improc}
\begin{figure}[ht]
    \hfuzz=5.002pt 
    \centering
    \includegraphics[width=\textwidth]{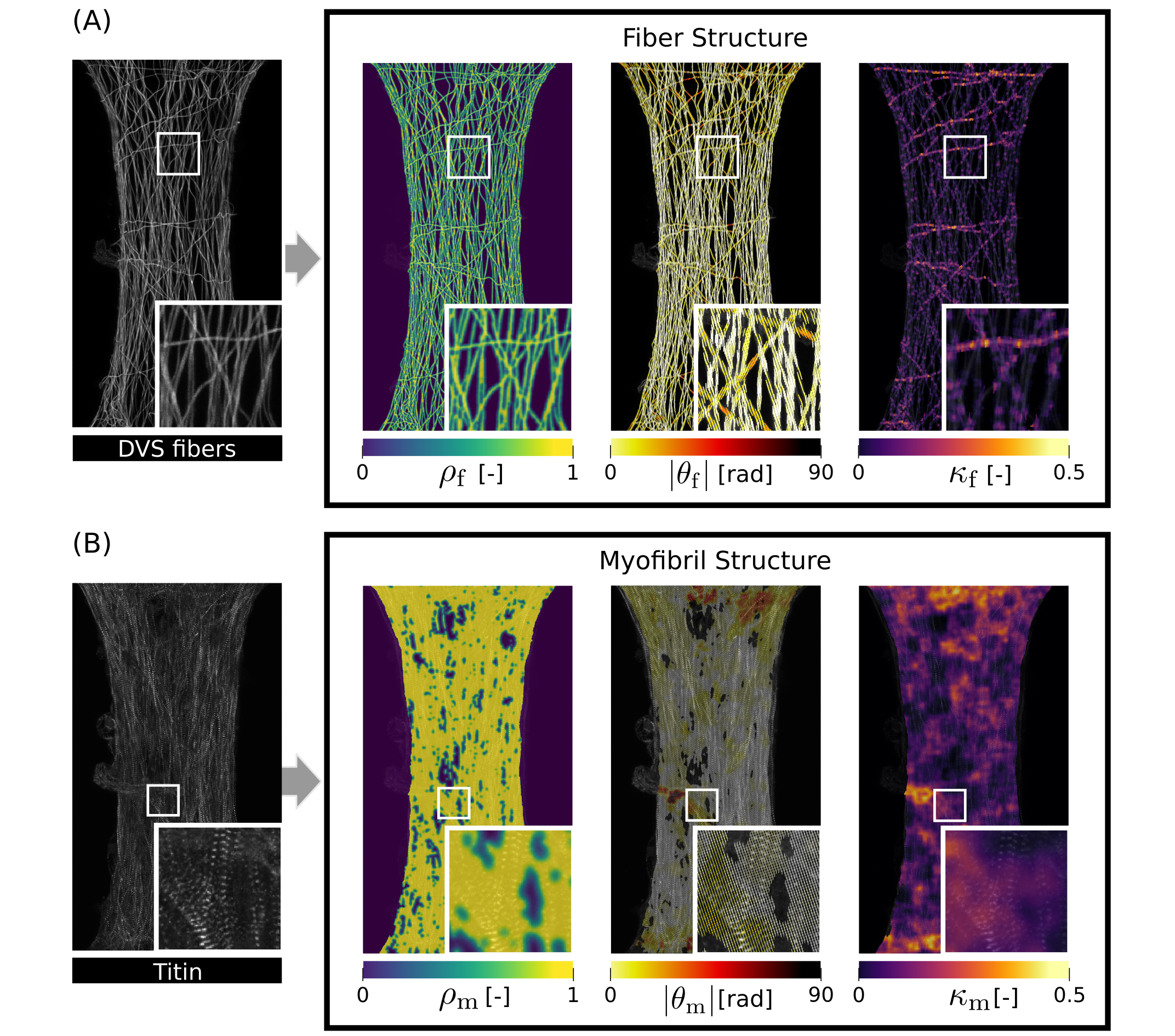}
    \caption{Image processing results for (A) the DVS fibers showing (from left to right) fiber density ($\rho_{\rm f}$), fiber alignment (shown as the absolute angle respect the post-to-post direction, $|\theta_{\rm f}|$), and fiber dispersion $\kappa_{\rm f}$, and for (B) For the titin images showing myofibril density $\rho_{\rm m}$, myofibril alignment (as the absolute angle respect the post-to-post direction, $|\theta_{\rm m}|$), and myofibril dispersion $\kappa_{\rm m}$.
    }
    \label{fig:improc}
\end{figure}

This subsection details the process of extracting information from the DVS fiber and titin images (Fig. \ref{fig:improc}) to obtain quantities that characterize the structure of the matrix and  the myofibril network that can then be projected into a 2D fibroTUG model.

\textit{DVS fibers}. The processing starts by creating a mask of the fibers that is used to compute the local fiber density $\rho_{\rm f}$, alignment $\mathbf f_0$, and dispersion $\kappa_{\rm{f}}$ as shown in Fig. \ref{fig:improc}A. The methods used to compute these quantities are detailed in Supplementary Information Section S1. The fiber density $\rho_f$ takes values between 0 (no fibers) and 1 (fiber), allowing us to define the mechanical presence of fibers. The fiber alignment vector, $\mathbf f_0$, provides the direction of the local stiffness anisotropy in the direction of the fibers. Finally, the dispersion $\kappa_{\rm f}$ is a parameter used in continuum models \cite{Schriefl2012} to represent regions where fibers follow a distribution around a mean vector. In our case, it allows the local stiffness to move from anisotropic (no dispersion, $\kappa_{\rm{f}}=0$) to isotropic ($\kappa_{\rm{f}}=0.5$) depending on the local fibers. 

\textit{Titin}. Myofibrils were identified via immunofluorescent imaging of iPSC-CMs containing a titin-GFP reporter, which allows for the visualization of the sarcomeres' Z-discs \cite{Sharma2018}. When images of titin were available for the whole domain, tissue-specific fields describing the structure of the myofibril network were computed. This was done following a similar strategy to the processing of the DVS fibers, where the titin images were used to find the myofibril density $\rho_{\rm m}$, alignment $\mathbf m_0$, and dispersion $\mathbf \kappa_{\rm m}$ (Fig. \ref{fig:improc}B). The steps to compute these quantities from the images are presented in Supplementary Information Section S2. The myofibril density $\rho_{\rm m}$ (taking values from 0 to 1) allows us to define the contractile regions. The alignment vector $\mathbf m_0$ defines the direction of the contraction and the dispersion $\kappa_{\rm m}$ activates an isotropic contraction in regions of unorganized myofibrils.

\begin{figure}[ht]
    \hfuzz=5.002pt 
    \centering
    \includegraphics[width=\textwidth]{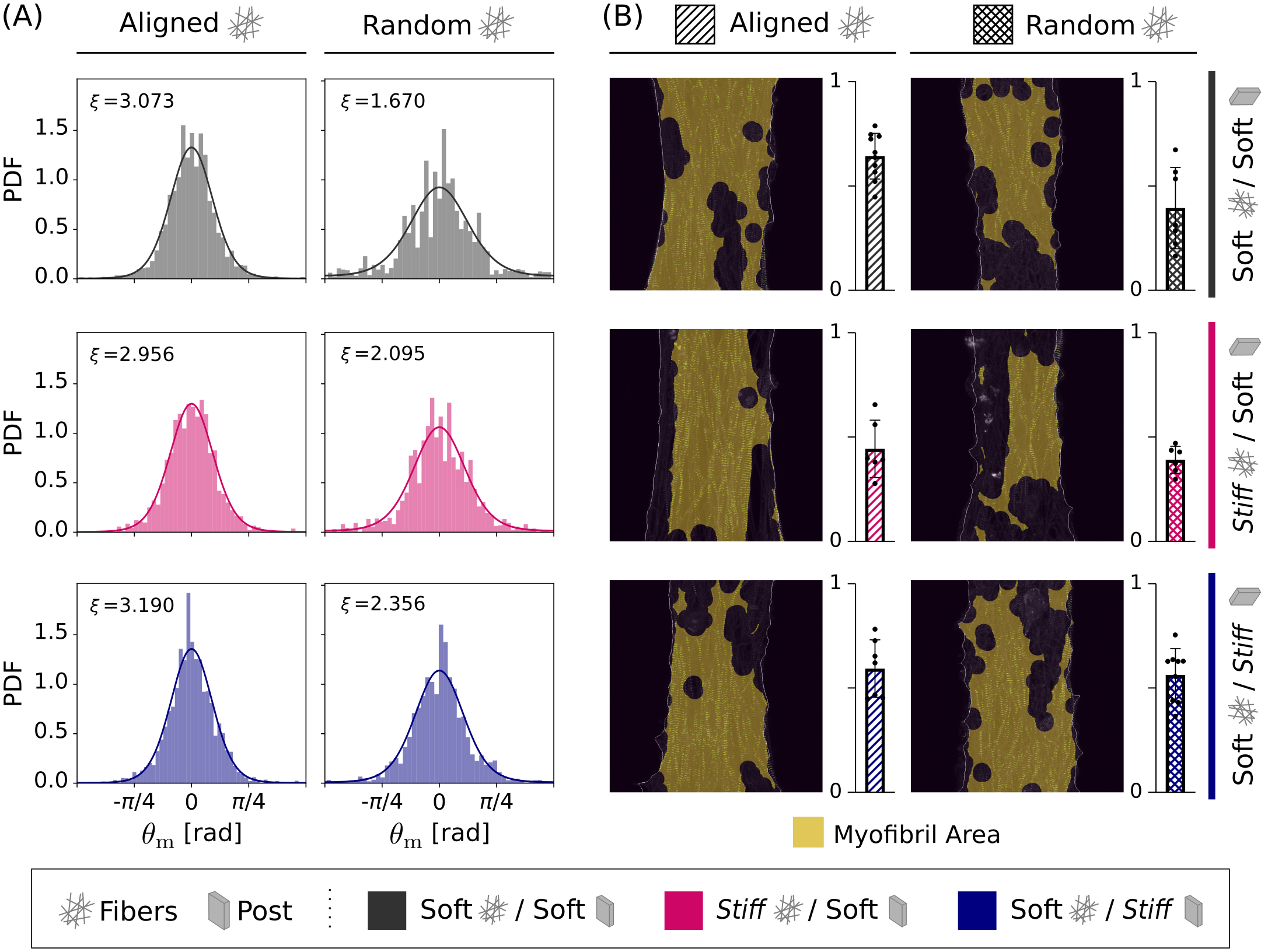}
    \caption{Myofibril characterization. (A) Histograms and their respective Von Mises fit for each condition studied. (B) Example image of the density computation for the different cases. The bar plots at the right show the density value computed for all the images available. All data presented as mean ± standard deviation.}
    \label{fig:myo_characterization}
\end{figure}

To obtain the visualization of the titin in the whole tissue shown in Fig. \ref{fig:expsetup}D, several smaller images were stitched together since a high magnification is needed to have a clear view of the Z-discs. This is a slow process, and since the main objective was to quantify myofibril structural characteristics, we decided to accelerate the process by only imaging the center of the tissues and statistically characterize the myofibril alignment and density. Given the six different experimental conditions, the myofibril orientation was analyzed and compiled for all the images available ($N\ge7$ per condition). A Von Mises distribution \cite{Schriefl2012} was fit to the histogram of angles (measured relative to the post-to-post direction, see Supplementary Information Section S2.1 for more details). The distribution is characterized by a parameter $\xi$ (assuming the mean is zero), with high values of $\xi$ indicating myofibrils oriented preferentially in the post-to-post direction. The resulting probability density functions (PDFs) and the data histograms are shown in Fig. \ref{fig:myo_characterization}A. The myofibril density was computed for these partial images and then a mean value was computed per condition (see Fig. \ref{fig:myo_characterization}B). The mean density values were then normalized by the aligned, soft fibers/soft post condition value. Table \ref{tab:myorho} shows the final density values used. Given a fiber network and geometry, we used the probability fitting and the density measurements specific for each condition to generate computational myofibril fields that followed these two parameters on top of the fiber fields. The procedure to create these fields is shown in Supplementary Information Section S3. One important thing to notice is that whenever this approach was taken, no myofibril dispersion was considered ($\kappa_{\rm m}=0$), as it is difficult to generate computationally in a meaningful way. The impact of these considerations was studied in Section S3.1 and S3.2 of the Supplementary Information.

\subsection{Biomechanical Model}

To model microtissue mechanics, we used a constrained mixture continuum mechanics framework \cite{Humphfrey2002,Holzapfel2000,Bonet2008}. Due to the thinness of the fibroTUG tissues ($\sim 12 \mu m$ compared to the $\sim 400 \mu m$ in length), we consider the tissue domain $\Omega \subset \R^2$. The boundary at the post is denoted by $\Gamma_p$. The reference coordinates of a given point in $\Omega$ are denoted by $\mathbf X$. Under internal and external loads, this point moves to a deformed position $\mathbf x = \mathbf X + \mathbf u$, where $\mathbf u$ is the displacement field. The deformation gradient tensor $\mathbf F = \nabla \mathbf u + \mathbf I$ describes the deformation of the material with respect to the reference coordinates and $J=\det{\mathbf F}$ the relative volume change \cite{Holzapfel2000}. We further define $\mathbf C = \mathbf F^T\mathbf F$ as the right Green Cauchy deformation tensors. Constitutive relations are often expressed in terms of the invariants of $\mathbf C$ \cite{Holzapfel2009}. In this work, we considered the following,
\begin{equation*}
	I_1=\mathbf{C}:\mathbf{I}, \qquad \bar{I}_1=J^{-2/3}I_1,
\end{equation*}
where $\bar{I}_1$ is the isochoric version of $I_1$ \cite{Nolan2014}. Furthermore, invariants describing the deformation in the fiber and myofibril direction are,
\begin{equation*}
\qquad I_{\rm 4f} =\mathbf{C}:\mathbf{f}_0\otimes\mathbf{f}_0, \qquad I_{\rm 4m} =\mathbf{C}:\mathbf{m}_0\otimes\mathbf{m}_0, \label{eq:C_invariants}
\end{equation*}
and these are further modified to account for local dispersion $\kappa_{\rm f}$, $\kappa_{\rm m}$ \cite{Schriefl2012, Holzapfel2019},
\begin{align*}
	I_{\rm 4f}^* = \kappa_{\rm f} I_1 + (1-2\kappa_{\rm f}) I_{\rm 4f}, \qquad
   I_{\rm 4m}^* = \kappa_{\rm m} I_1 + (1-2\kappa_{\rm m}) I_{\rm 4m}.
\end{align*}
When the fiber dispersion $\kappa_{\rm f}=0$, $I_{4\rm f}^*=I_{4\rm f}$, i.e., the local response of $I_{4\rm f}^*$ is fully anisotropic, whereas when $\kappa_{\rm f}=0.5$,  $I_{4\rm f}^*=I_{1}/2$, and the material behaves locally as a fully isotropic material. The same is true for the myofibril invariant $I_{4\rm m}^*$.

As mentioned in Section \ref{sec:methods:expsetup}, a fibroTUG tissue consists of two components, DVS fibers and iPSC-CMs. We assumed the two components work in parallel and, therefore, the strain energy density of the tissue $\Psi$ is given by the sum of the energy of the fibers $\Psi_{ECM}$ and the iPSC-CMs $\Psi_{CM}$,
\begin{equation}
	\Psi = \Psi_{ECM} + \Psi_{CM},
\end{equation}

The strain energy function $\Psi_{\rm ECM}$ is given by the strain energy density describing the fiber mechanics $\Psi_{\rm f}$ and a term that delivers numerical stability on the areas where there are no fibers $\Psi_{\rm st}$,
\begin{align*}
	\Psi_{ECM} = \Psi_{\rm f}(\mathbf C; \rho_{\rm f}, \mathbf f_0, \kappa_{\rm f}) + \Psi_{\rm st}(\mathbf C).
\end{align*}
The first term corresponds to a modified neofiber material law \cite{Hadjicharalambous2014} and integrates the structural information obtained from the DVS fiber images,
\begin{align} 
	\Psi_{\rm f}(\mathbf C; \rho_{\rm f}, \mathbf f_0, \kappa_{\rm f}) = \rho_{\rm f}  \left( \frac{C_1}{4} (I_{4\rm f}^*-1)^2 + \frac{C_2}{2}(\bar I_1-2) \right), \label{eq:Psi_f}
\end{align}
where $C_1$ is the stiffness in the fiber direction and $C_2$ is the isotropic stiffness. This formulation was chosen since the experimental data from the passive stretching of the fibers (Fig. \ref{fig:expsetup}B) showed a close-to-linear behavior that is well captured by this material law. The stabilization term is,
\begin{equation}
    \Psi_{\rm st} = \frac{K}{2}\left[(J-1)^2  + (\ln J)^2 \right] + \mu \tr (\mathbf E^2).
\end{equation}
The first term inside the parenthesis penalizes volumetric changes, while the second corresponds to the deviatoric term of a Saint-Venant Kirchhoff material. The parameters $K=10^{-3}$ kPa and $\mu=10^{-2}$ kPa are chosen to be $\ll C_1$, so the mechanical response of the ECM is dominated by $\Psi_{\rm f}$. A sensitivity analysis confirmed that these parameters are minimally important (see Section S6 of the Supplementary Information for more details).

The contractile iPSC-CMs were modeled by a passive component representing the bulk stiffness of the cells ($\Psi_{\rm c}$) and an active contraction component ($\Psi_{\rm a}$),
\begin{equation}
	\Psi_{\rm CM} = \Psi_{\rm c} + \Psi_{\rm a}.
\end{equation}
The passive component is modeled using the 2D version of the compressible Neohookean material law presented in Pascon et al. (2019) \cite{Pascon2019}, which has the following strain energy density function,
\begin{equation}
	\Psi_{\rm c} = \frac{K_c}{2}[\ln J]^2 + \frac{\mu_{\rm c}}{2} \left(I_1 - 2 - 2\ln J\right).
\end{equation}
We use $\mu_{\rm c}=2$ kPa, which is close to values derived from stress-strain curves obtained from isolated iPSC-CMs \cite{Ballan2020}. Due to the 2D nature of the EHTs, the compressible modulus was assumed negligible (i.e. $K_c=0$) due to the ease for cells to deform out-of-plane under in-plane compression.

The active component is given by,
\begin{equation}
\Psi_{a}  = \int_0^{I_{\rm 4m}^*} \eta \rho_{\rm m}\phi(s) \; {\rm d}s. \label{eq:Psi_a}
\end{equation}

Here, $\phi$ is a function that models the length-dependent behavior of cardiomyocytes, and it is taken from \cite{Kerckhoffs2003} (see Supplementary Information Section S4 for more details). The parameter $\eta$ controls the magnitude of the activation in time of the iPSC-CMs. Note that we assume the activation only occurs where there are myofibrils (hence the multiplication by $\rho_{\rm m}$), but the passive response of the iPSC-CMs, is present everywhere in the tissue. This is because iPSC-CMs nuclei appear evenly distributed in the images of the tissues, but only some develop myofibrils.

The total Cauchy stress $\bm \sigma$ is computed from $\Psi$ using,
\begin{equation}
    \bm\sigma = J^{-1}\pd{\Psi}{\mathbf F}\mathbf{F}^T,
\end{equation} 
and, under a quasi-static regime, the stress balance is given by,
\begin{equation}
    \nabla\cdot \bm\sigma = 0.
\end{equation}

Throughout the paper, we assess the mechanics of the tissues using the active stress and the strain in the myofibril direction,
\begin{equation}
    \sigma_{\rm a, m} = \bm \sigma_{\rm a}: \mathbf m\otimes \mathbf m, \qquad \varepsilon_{\rm m} = \frac12 \left(\mathbf C - \mathbf I \right): \mathbf m\otimes \mathbf m.
\end{equation}
where $\bm \sigma_{\rm a} = J^{-1}\pd{\Psi_{\rm a}}{\mathbf F}\mathbf{F}^T$.

\subsection{Data Assimilation} \label{sec:methods:assimilation}

Once the structure of each tissue is defined, the only unknowns remaining in the system are the material parameters of the constitutive law of the fibers, $C_1$ and $C_2$, and of the active component of the iPSC-CMs, $\eta$. For simplicity, we considered these parameters to be constant within a fibroTUG. To find these tissue-specific values, we assimilate the functional data shown in Fig. \ref{fig:expsetup}C-D using a parameter identification strategy introduced by \cite{Miller2021}. Briefly, this technique integrates material parameters into the set of state variables, enabling the addition of constraints to match measured data. This method allows solving for both the displacement field and additional material parameters in the same forward simulation. In our case, we divide the assimilation into two steps. First, we use the data from the fiber indentation test (Fig. \ref{fig:expsetup}C) to find the stiffness parameters of the fibers, and then we use the post-displacement trace (Fig. \ref{fig:expsetup}D) to find the active parameter, $\eta$. The corresponding boundary value problem equations are described in Section S5 of the Supplementary Information.

\section{Results}

\subsection{Model validation} 
\begin{figure}[ht]
    \hfuzz=5.002pt 
    \centering
    \includegraphics[width=\textwidth]{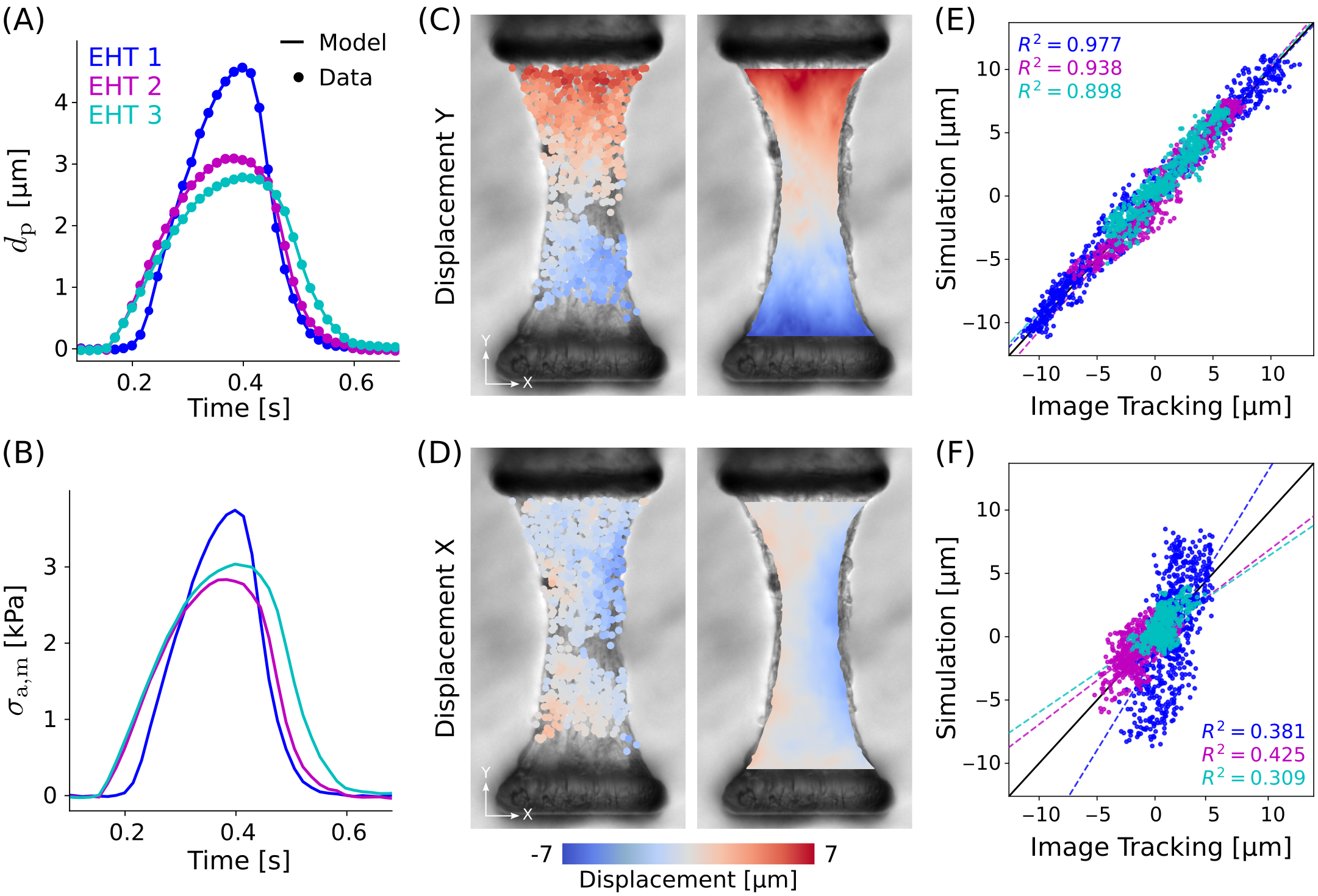}
    \caption{Results of the pipeline for three complete datasets. Each color represents a different tissue. (A) Post displacement data (dots) and model (line). (B) Simulated active stress. (C) MicroBundleCompute displacement outputs (left) and simulation results (right) in the post-to-post direction Y. (D) Measured displacements (left) and simulation results in the X direction. (E) Correlation between the tracking and the simulation displacements in Y and (F) in X. (C) and (D) correspond to the case pictured in cyan.}
    \label{fig:validation}
\end{figure}

The objective of this first set of simulations was to assess the ability of the proposed pipeline and model to capture fibroTUG mechanics. To do so, three complete datasets of the soft fibers/soft post condition were analyzed. The fiber stiffness was set to the mean value found for the fibrous samples of this condition, $C_1=3.84$ kPa (see Section \ref{sec:results:exp}). By design, the post-displacement (and, therefore, the post-force) data were exactly matched by the simulations as shown in Fig. \ref{fig:validation}A. The data assimilation enabled us to identify the parameter $\eta$ that reproduces the post-displacement and force curves. Fig. \ref{fig:validation}B shows the mean $\sigma_{\rm a,m}$ at each time point for each tissue. To assess the ability of the model to capture local deformations, we validated the results of the simulations against the results from MicroBundleCompute, an image tracking software developed specifically for measuring the internal displacements of contracting EHTs \cite{Kobeissi2023}. For the three tissues, correlation plots were calculated between the predicted displacements of the simulation and the measured displacements from the video. Fig. \ref{fig:validation}C and D show the measured and predicted displacement fields in the post-to-post direction (Y) and the perpendicular direction (X) for one of the three datasets. Fig. \ref{fig:validation}E and F show the displacement's correlation in Y and X, respectively. The $R^2$ parameter for the correlation in the Y direction is always greater than 0.89, while in the X direction, a positive correlation is observed, with $R^2$ values close to 0.4.

\subsection{Active stress prediction for experimental conditions} \label{sec:results:exp}
In this section, we focus on understanding the effect of different mechanical environments on the iPSC-CMs ability to exert contractile stress. Images of five fibrous matrices for each condition were processed as described in \ref{sec:methods:improc} (with the exception of the random, soft fibers/soft post, where only four matrices with the paired indentation test were available). The fiber stiffness for each available matrix was identified using the indentation experimental data as detailed in \ref{sec:methods:assimilation}. As expected, the mean fiber stiffness for the stiff matrix cases (mean $C_1=19.47$/$C_1=24.66$ for aligned/random matrices) is about four times stiffer than the soft matrix case (mean $C_1=3.84$/$C_1=4.26$ for aligned/random matrices). More detailed results of this process can be found in Supplementary Information Section S8. We computationally generated five myofibril fields on top of each matrix following the probabilistic characterization shown in Fig. \ref{fig:myo_characterization}. Further, five representative samples (i.e., that have a similar mean and standard deviation than the whole set) of post-displacement curves were used as input in the active data assimilation step. This gives us $N=125$ \textit{in-silico} models per experimental condition (except for the random, soft fibers/soft post condition where $N=100$). 

\begin{figure}[ht]
    \hfuzz=5.002pt 
    \centering
    \includegraphics[width=\textwidth]{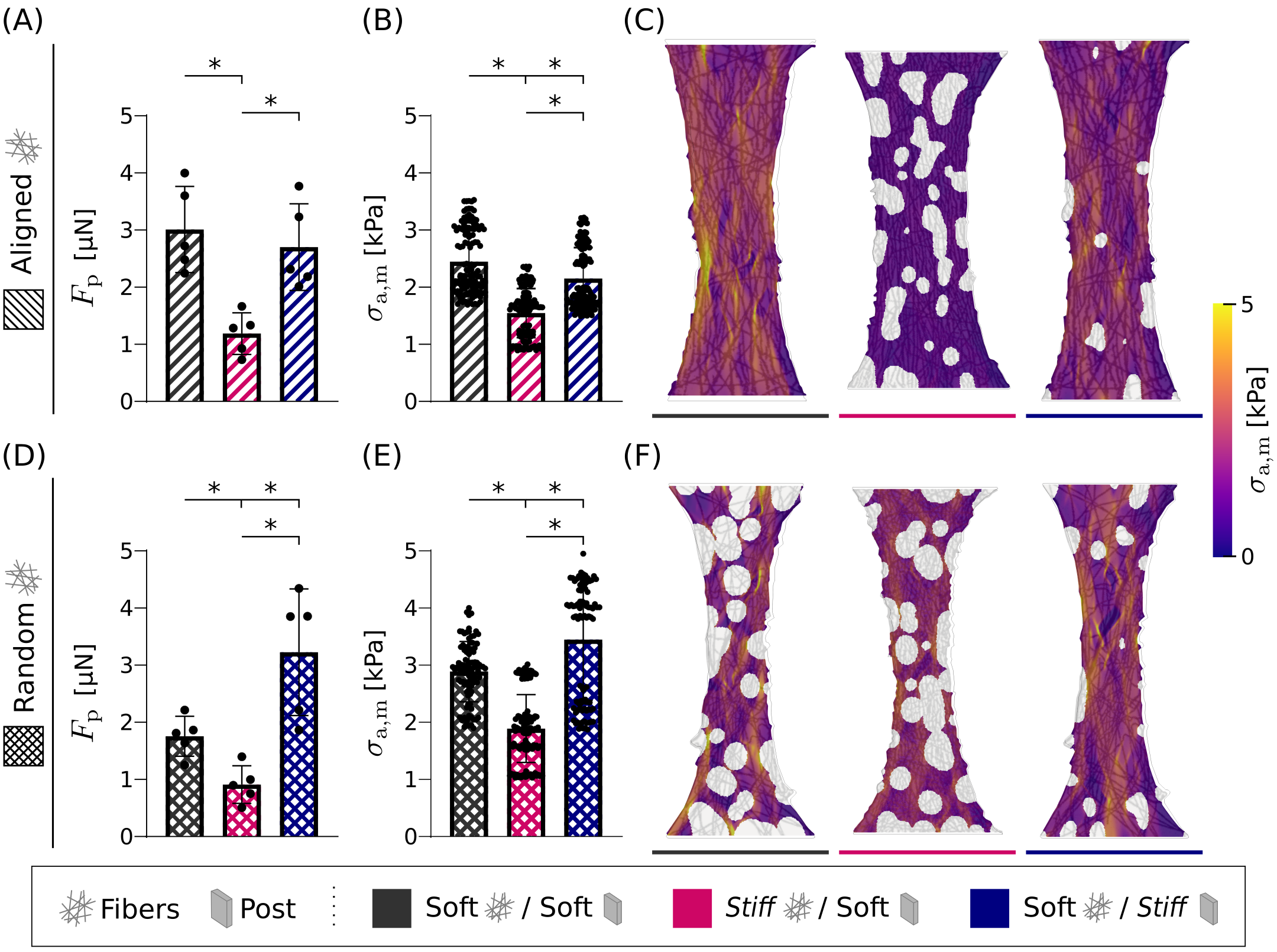}
    \caption{Results at maximum contraction for the experimental conditions of interest. (A) Force output, (B) mean $\sigma_{\rm a,m}$, and (C) $\sigma_{\rm a,m}$ field at maximum contraction for aligned fibrous matrix. (D) Force output, (E) mean $\sigma_{\rm a,m}$, and (F) $\sigma_{\rm a,m}$ field at maximum contraction for random fibrous matrix. All data presented as mean $\pm$ standard deviation. * $p<0.0001$ by unpaired t-tests.}
    \label{fig:expresults}
\end{figure}

Fig. \ref{fig:expresults}A-C shows the results for aligned matrices, including bar plots for the post-force and the mean active stress $\sigma_{\rm a, m}$, and the $\sigma_{\rm a, m}$ field at maximum contraction for aligned matrices. Fig. \ref{fig:expresults}D-F shows the same for random matrices. Both $F_{\rm p}$ and $\sigma_{\rm a,m}$ follow the same trends when comparing condition to condition. For instance, in the aligned case, the magnitude of both $F_{\rm p}$ and $\sigma_{\rm a,m}$ are highest in the soft fibers/soft post-condition and lowest in the case with stiff fibers. However, the relative differences between conditions in $\sigma_{\rm a,m}$ are lower compared to $F_{\rm p}$. For example, the force output drops from $3.01\, \mu$N in the soft fibers/soft post case (grey bars in \ref{fig:expresults}A) to only $1.19\, \mu$N in the case with stiffer fibers (red bars in \ref{fig:expresults}A), which corresponds to a -60.5\% relative change. Contrarily, the maximum $\sigma_{\rm a,m}$ falls from 2.45 to 1.55 kPa between these two cases, only a -36.8\% variation.

\subsection{Isolating the effect of mechanical variations}

\begin{figure}[ht]
    \hfuzz=5.002pt 
    \centering
    \includegraphics[width=0.9\textwidth]{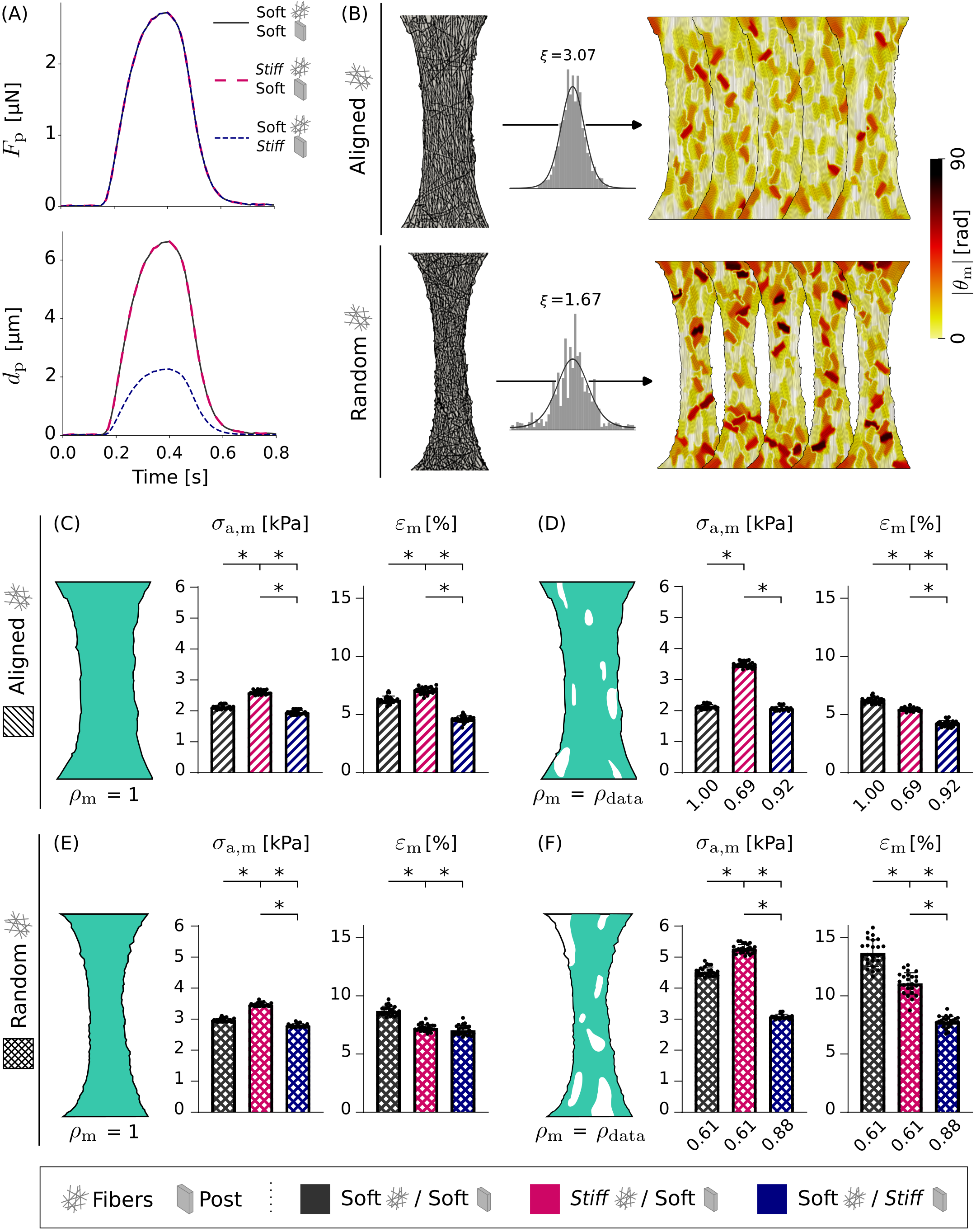}
    \caption{Impact of different mechanical conditions on the post-force/active stress relationship. (A) Post-force and post-displacement for each condition. Note that the force is kept the same for all conditions, while the post-displacement varies in the case of stiff posts. (B) Procedure to generate \textit{in-silico} samples (see supplementary Information S3 for more details). An aligned and a random matrix were built from the images, and computationally myofibril fields were created using the corresponding PDF. Results for an aligned condition (C) with  $\rho_{\rm m}=1$, and (D) with $\rho_{\rm m}$ equal to the specific density obtained from data for each case ($\rho_{\rm data}$). (E) and (F) show the results for random matrices with $\rho_{\rm m}=1$ and $\rho_{\rm m}=\rho_{\rm data}$. The values below the bar plots in (D) and (F) correspond to the mean density used in these cases (see Table \ref{tab:myorho}). All data presented as mean $\pm$ standard deviation. * $p<0.0001$ by unpaired t-tests.}
    \label{fig:mm}
\end{figure}

To investigate the origin of the $F_{\rm p}/\sigma_{\rm a, m}$ relative differences, we performed several simulations outside the parameter range of current experimental conditions. Specifically, we studied variations in post stiffness, fiber stiffness, fiber alignment, and myofibril density. To do so, we took a single post-force curve (Fig. \ref{fig:mm}A) and computed the active stress needed to generate that force output given a certain parameter set. In this experiment, higher values of $\sigma_{\rm a,m}$ indicate that the tissue is less efficient in transmitting cell stress to force output as this was held constant across all conditions. We performed this test using five random matrices and five aligned matrices with five computationally generated myofibril fields each (Fig. \ref{fig:mm}B). The alignment of myofibrils was taken from the soft post/soft fibers case (gray curves in Fig. \ref{fig:myo_characterization}A). The stiffness in the fiber direction was set to be $C_1=3.84$ and $C_1=19.47$ kPa, for the soft fiber and the stiff fiber case, respectively. 

First, we considered a constant and uniform myofibril density $\rho_{\rm m}=1$. Results of this experiment are shown in Fig. \ref{fig:mm}C, for aligned matrices and Fig. \ref{fig:mm}E, for random matrices. On aligned matrices, the active stress necessary to generate the force output represented in Fig. \ref{fig:mm}A was $2.12,2.60,1.95$ kPa for the soft fibers/soft post, stiff fibers, and stiff post conditions, respectively. Similar relative trends are observed for the random case, although with higher values compared to their aligned counterparts. Second, to study the effect of myofibril density, we considered the case where $\rho_{\rm m}=\rho_{\rm data}$, which is the specific density computed from the data (shown in Table \ref{tab:myorho}). Fig. \ref{fig:mm}D shows the results of these simulations. The magnitude of $\sigma_{\rm a, m}$ is elevated in those cases with lower $\rho_{\rm m}$, when compared with the $\rho_{\rm m} = 1$ simulations for both aligned and random matrices. 

\begin{figure}[ht]
    \hfuzz=5.002pt 
    \centering
    \includegraphics[width=\textwidth]{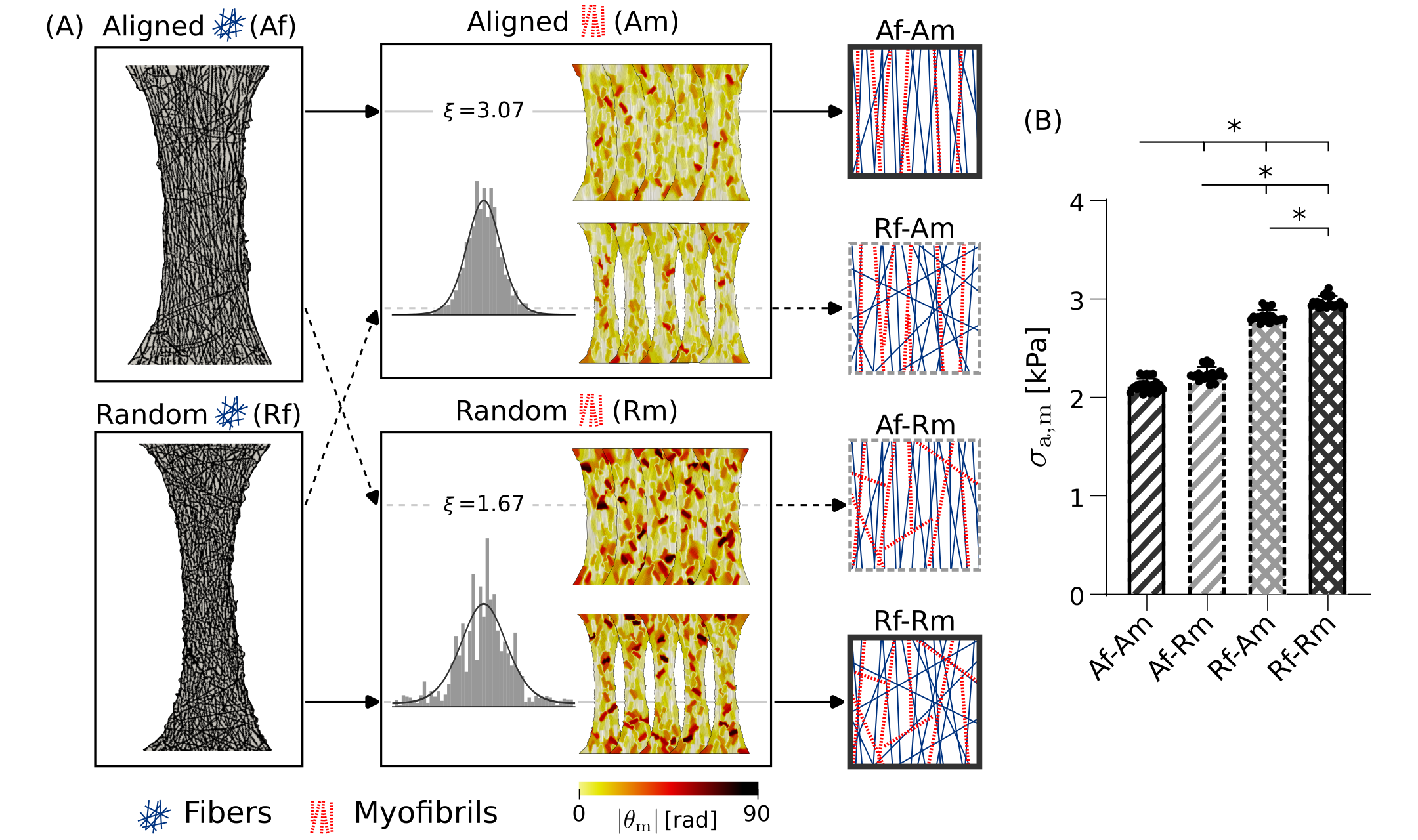}
    \caption{Procedure for the myofibril vs fiber alignment test. Simulated conditions that have aligned fibers with random myofibrils (Af-Rm) and vice-versa (Rf-Am) \textit{in-silico} tissues are generated by using fibrous matrices generated from images and the PDFs describing myofibril alignment. These simulations are compared with the aligned (Af-Am) and random (Rf-Rm) experimental conditions. (B) Mean $\sigma_{\rm a,m}$ at maximum contraction for the Af-Am, Af-Rm, Rf-Am, Rf-Rm, simulations using soft fibers/soft posts. All data presented as mean $\pm$ standard deviation. * $p<0.0001$ by unpaired t-tests.}
    \label{fig:mm_mf}
\end{figure}

Finally, to understand the reason behind the performance of random matrices compared to aligned matrices, we investigated the relative importance of myofibril alignment and fiber alignment in the transmission of iPSC-CM stress to post-force. To do so, we simulated mixed, synthetic scenarios where aligned myofibril fields ($\xi=3.07$) were generated on top of random fiber fields (Rf-Am) and vice-versa, random myofibrils ($\xi=1.67$) on aligned fibers (Af-Rm) as shown in Fig. \ref{fig:mm_mf}A. These cases were compared with the experimental scenarios, aligned fibers with aligned myofibrils (Af-Am) and random fibers with random myofibrils (Rf-Rm), which correspond to the bar plots shown in Fig. \ref{fig:mm}C,E. We performed the same experiment as in the previous section with these two mixed cases. Results for the soft fibers/soft post case are shown in Fig. \ref{fig:mm_mf}B. The results for the other conditions (with stiffer fibers or stiffer posts) are very similar and are shown in Supplementary Information Section S9. The difference in $\sigma_{\rm a,m}$ between the experimental conditions Af-Am and Rf-Rm is 0.846 kPa. When random myofibrils are used on top of aligned fibers (Af-Rm), that difference is reduced to 0.731 kPa (-13.6\%). When myofibrils are aligned in random matrices (Rf-Am) $\sigma_{\rm a,m}$ the reduction is similar (-16.7\%). The highest variation is produced by changing the matrix alignment. For example, the difference is reduced by 86.4\% when we fixed the myofibril alignment to be aligned and changed the matrix alignment (i.e., from Af-Am to Rf-Am).

\section{Discussion}

In this work, we developed a pipeline to generate data-driven computational models of EHTs. By combining comprehensive experimental data with biomechanical computational models, we are able to model EHT mechanics and compute metrics to estimate iPSC-CM function. The pipeline developed allows us to model the explicit fiber structure and myofibril network and to infer the mechanical properties of these components from functional experimental data. This is facilitated by the use of the fibroTUG platform, which provides great control over the mechanical environment and enables obtaining detailed imaging and functional data \cite{DePalma2023}. The biomechanical computational model augments the analysis and conclusions obtained from the experiments. It also enables the test of conditions that are experimentally not feasible to achieve, allowing us to decipher the effect of the different variables of the system in the tissue mechanics. Below, we discuss the results of the different simulations performed using the model.

\subsection{Model Validation}

For the three validation datasets, the correlation between the displacements predicted by the simulations and the measurements of the MicroBundleCompute software in the post-to-post direction is strong, with high $R^2$ values (mean 0.938) and slopes close to 1 (0.967, 1.047, 0.970 for the three tissues, see Fig. \ref{fig:validation}E). The correlations in the perpendicular direction are also positive but with slopes further from 1 (1.495, 0.694, 0.623 for the three tissues) and a mean $R^2=0.372$ (see Fig. \ref{fig:validation}F). The decrease in correlation in the X directions is explained by several reasons. First and foremost, the videos are taken using brightfield imaging \cite{DePalma2023}, which mainly shows the fiber's deformations. Our modeling approach uses a constrained mixture framework, where the kinematics of fibers and cells are homogenized, meaning that the computed displacements represent the average displacement of both these components. Furthermore, the displacements in the X direction are smaller and harder to track as demonstrated by a simple exercise where two people track different features across the contraction cycle. This test showed a decrease in the measured displacement correlation, from $R^2=0.9$ in the Y direction to $R^2=0.72$ in the X direction (see Supplementary Information Section S7 for more details). For these reasons, we believe that imaging-to-modeling comparison is not direct, but it does allow us to confirm that our model is meaningfully capturing the main features of fibroTUG kinematics.

\subsection{Active stress prediction for experimental conditions}
In Fig. \ref{fig:expresults}, the computational pipeline was used to assess the mechanics of fibroTUGs under different mechanical environments. Both $F_{\rm p}$ and $\sigma_{\rm a,m}$ show significant differences between the conditions (Fig. \ref{fig:expresults}A-B,D-E). Since $\sigma_{\rm a,m}$ is a value that considers the structure of all the tissue components, the differences in this value indicate that modifying the mechanical environment where the cells develop will influence their contractile maturity. These differences are also captured by $F_{\rm p}$, but the relative differences are influenced by a multitude of factors, including fiber mechanics, myofibrillar density and alignment, and contractile maturity. This proves that the force output is the reflection of different variables in the system, not only the iPSC-CMs active stress. This highlights the importance of giving proper context to force output, which is often treated as a direct surrogate of iPSC-CM stress \cite{RonaldsonBouchard2018,Leonard2018,Boudou2012}.

\subsection{Isolating the effect of mechanical variations}
One key advantage of using computational models to study EHT mechanics is their flexibility to study scenarios that are difficult, if not impossible, to create \textit{in-vitro}. These simulations can shed light on the influence of different variables of a system by allowing us to change their values individually and measure the variation in an output of interest. We performed controlled simulations where one mechanical variable was modified at a time to assess the changes in the $F_{\rm p}$/$\sigma_{\rm a,m}$ relationship. In Fig. \ref{fig:mm}C and E, we can see that when $\rho_{\rm m}=1$, changing the ECM stiffness and the boundary conditions will have a slight impact on the capacity of the tissue to translate cell active stress into force output. For example, for stiffer fibers, a higher amount of the work done by the iPSC-CMs is lost in tugging the less deformable matrix. Conversely, when the post is stiffer, the magnitude of $\sigma_{\rm a, m}$ is slightly lower, corresponding to a more efficient transduction of myofibrillar stress to the total output force.

When $\rho_{\rm m}=\rho_{\rm data}$, the conditions with lower myofibril formation need to generate higher values of $\sigma_{\rm a, m}$. This is not surprising because there will be less myofibril area to generate the same force, which means that the existing sarcomeres need to generate more stress to compensate. However, our model allows us to measure that effect and compare it with the effect of other mechanical variations. The observed changes due to poor myofibril formation are much higher than those observed due to post or fiber stiffness (observed by the change of the bar plot magnitude between Fig. \ref{fig:mm}C-E and Fig. \ref{fig:mm}D-F). Furthermore, the cases that have lower $\rho_{\rm m}$ (and that need higher $\sigma_{\rm a, m}$ to generate the same force) correlate with the cases that have lower force output experimentally (Fig. \ref{fig:expresults}A-B). This indicates that myofibril density is one of the most important mechanical parameters explaining EHT force output. The importance of myofibril formation has also been observed in single-cell models \cite{Roest2021}.

When evaluating the effect of fiber alignment, we can see that the magnitude of $\sigma_{\rm a, m}$ is higher in random than aligned matrices. This makes sense, as in these cases, both fibers and myofibrils are less aligned in the post-to-post direction (see, for example, the PDFs in Fig. \ref{fig:myo_characterization}A). Here, a lot of the work performed by the iPSC-CMs gets lost in pulling the transverse direction. In Fig. \ref{fig:mm_mf}, we decouple the alignment of fibers and myofibrils with the intention of understanding the relative importance of myofibril alignment versus fiber alignment. The results show that aligning the myofibrils on top of a random matrix (Am-Rf) or vice-versa (Rm-Af) only explains a small part of the difference between the all-aligned (Af-Am) and the all-random condition (Rf-Rm). For this reason, we conclude that, \emph{for the range of observed variations in myofibril alignment}, the matrix structure has a higher effect on the force output. It could be that other mechanical environments generate a higher disarray of myofibrils making this parameter more dominant. 

The results of the model complement nicely with other biomarkers studied in this platform \cite{DePalma2023}. For example, the aligned soft fibers/soft post condition showed a higher proportion of connexin 43, more mature forms of myosin (MLC-2v), and lower beats per minute, to mention a few. Our model shows that this condition and the aligned soft fibers/stiff post are the most efficient in transmitting force. However, the case with soft posts presents much higher myofibril strains (see Fig. \ref{fig:mm}C), which has shown to be important in iPSC-CM maturation \cite{Tsan2021}. These observations show the benefits of using a combined computational-experimental approach to understand different aspects that are involved in the EHT function and how they relate to the iPSC-CM maturation.

\subsection{Limitations}

The computational generation of myofibril fields enabled the flexibility to explore non-experimental scenarios to understand the importance of myofibril alignment. However, generating myofibrils networks that are representative of real sarcomere organization is challenging. We performed several tests to assess our approach using the validation datasets. We tested the importance of including myofibril dispersion $\kappa_{\rm m}$ and studied the prediction error induced when using computationally generated myofibril fields instead of imaged-based ones. Detailed results of these tests are shown in Supplementary Information Section S3. Briefly, we showed that not including $\kappa_{\rm m}$ will mainly impact the deformations in X, with these simulations showing less necking than expected. However, the active stress prediction remains very similar. The test of the artificial myofibril fields showed that the sarcomere strain values are the most affected quantity, but the active stress prediction is, again, very similar to the one computed using image-generated myofibril fields. The results of these tests show that we can use our proposed method to correctly assess changes in $\sigma_{\rm a,m}$. 

A few assumptions were made in the model not based on the experimental data. The passive stiffness of the iPSC-CMs, isolated from the fibers, was not measured in our experiments. This parameter is tricky to measure, as it is known that the cell will change its stiffness during development \cite{Lahmers2004} and that this value also depends on the temperature and the state of its contractile apparatus \cite{King2010}. Therefore, for simplicity, we considered the passive stiffness of the cells to be 2 kPa, a value similar to the one measured by Ballan et al. \cite{Ballan2020} in iPSC-CMs, and lower than what is usually measured in mature CMs \cite{Caporizzo2020}. We further assume that the passive response of the cell was isotropic. This is because no studies were found on the anisotropy of iPSC-CMs and since the iPSC-CMs considered in this study are still immature (thus, the cytoskeleton is less organized), we believe that our assumptions of an isotropic, less stiff passive response than in adult CMs are reasonable. 

Another modeling assumption is that all iPSC-CMs have a synchronous activation and that all their contraction can be modeled using a single $\eta$ parameter. This means that the spatial heterogeneity observed in the $\sigma_{\rm a,m}$ fields in Fig. \ref{fig:expresults}C,F are the product of only the length-dependent function $\phi$ in Eq. \eqref{eq:Psi_a}. The simultaneous activation assumption is backed up by these tissues being very small, and the calcium transients measured experimentally did not show spatial differences \cite{DePalma2023}. However, it is possible (and probably expected) that not all the iPSC-CMs in one tissue have the same maturity, which could be modeled by having an $\eta$ field instead of a single scalar value. For example, instead of only using the post-displacement/force as a constraint, we could use the displacement measurements from the MicroBundleCompute software to find a local $\eta$. Different strategies to do so will be explored in the future. Nevertheless, we believe that using a single $\eta$ parameter allows us to compute an average activation and that the overall regional differences will not affect the conclusions of this paper.  

Finally, we modeled the fibroTUG in 2D, which also forces us to consider a material with low compressibility. This decision is based on the low thickness to length ratio that is $\sim 3\%$, and because only 2D projected images were available for the fibers, myofibrils, and displacement tracking. Future studies will aim to generate a pipeline to reconstruct the full tissue geometry from 3D stacks obtained from high-magnification imaging techniques and assess the validity of the 2D model.

\section{Conclusions}

This work introduces a data-driven biomechanical computational model of EHTs. The implemented pipeline leverages the experimental data of the fibroTUG platform, which was introduced specifically to study iPSC-CMs function due to mechanics. With the model, we are able to measure the iPSC-CM active stress that replicates the experimental observations. Unlike the tissue force output, this value is a direct measure of iPSC-CM contractile function. We used the model to assess the effects of different mechanical environments on iPSC-CMs stress generation. The results followed the same trends as the force output, indicating maturation differences, but with lower relative differences. This highlights that EHT force generation is the product of different variables besides iPSC-CM contraction. We explore this fact with our model and show that myofibril density is one of the key factors explaining the experimental differences observed. The developed framework opens the door to many other applications in the cardiac tissue engineering field, and it shows how a combination of \textit{in-silico} with \textit{in-vitro} approaches can help us better understand the mechanobiology of EHTs.

\section{Acknowledgements}

JJ acknowledges the support of ANID and Fulbright Chile through the Becas BIO program. DN acknowledges funding from the Engineering and Physical Sciences Research Council Healthcare Technology Challenge Award (EP/R003866/1). BMB acknowledges the support from the National Science Foundation through CBET-2033654. All authors acknowledge the support from the National Science Foundation through the Nanosystems Engineering Research Center for Directed Multiscale Assembly of Cellular Metamaterials with Nanoscale Precision (CELL-MET, EEC-1647837).

\newpage
\begin{table}[ht]
\centering
\caption{Results of the myofibril density characterization (mean $\pm$ standard deviation) for the different experimental conditions. The normalized density is the mean density of each case divided by the aligned soft fibers/soft post condition mean density (i.e., divided by 0.643).}
\label{tab:myorho}
\begin{tabular}{l|rr||rr|}
\cline{2-5}
 & \multicolumn{2}{c||}{Aligned Fibers} & \multicolumn{2}{c|}{Random Fibers} \\ \hline
\multicolumn{1}{|l|}{Condition} & \multicolumn{1}{c|}{Density} & \multicolumn{1}{c||}{Normalized Density} & \multicolumn{1}{c|}{Density} & \multicolumn{1}{c|}{Normalized Density} \\ \hline
\multicolumn{1}{|l|}{Soft Fibers / Soft Post} & \multicolumn{1}{r|}{0.643 $\pm$ 0.104} & 1.000 & \multicolumn{1}{r|}{0.395 $\pm$ 0.180} & 0.613 \\ \hline
\multicolumn{1}{|l|}{Stiff Fibers / Soft Post} & \multicolumn{1}{r|}{0.444 $\pm$ 0.126} & 0.690 & \multicolumn{1}{r|}{0.391 $\pm$ 0.060} & 0.608 \\ \hline
\multicolumn{1}{|l|}{Soft Fibers / Stiff Post} & \multicolumn{1}{r|}{0.593 $\pm$ 0.128} & 0.922 & \multicolumn{1}{r|}{
0.563 $\pm$ 0.119} & 0.875 \\ \hline
\end{tabular}
\end{table}

\bibliography{bibliography}
\end{document}